# Spliced Leader Trapping Reveals Widespread Alternative Splicing Patterns in the Highly Dynamic Transcriptome of *Trypanosoma brucei*

Daniel Nilsson[1], Kapila Gunasekera[1], Jan Mani[1], Magne Osteras[2], Laurent Farinelli[2], Loic Baerlocher[2], Isabel Roditi[1]*, Torsten Ochsenreiter[1]*

1 Institute of Cell Biology, University of Bern, Bern, Switzerland, 2 Fasteris, Genome Analyzer Service FASTERIS SA, Geneva, Switzerland

## Abstract

*Trans*-splicing of leader sequences onto the 5′ends of mRNAs is a widespread phenomenon in protozoa, nematodes and some chordates. Using parallel sequencing we have developed a method to simultaneously map 5′splice sites and analyze the corresponding gene expression profile, that we term spliced leader trapping (SLT). The method can be applied to any organism with a sequenced genome and *trans*-splicing of a conserved leader sequence. We analyzed the expression profiles and splicing patterns of bloodstream and insect forms of the parasite *Trypanosoma brucei*. We detected the 5′ splice sites of 85% of the annotated protein-coding genes and, contrary to previous reports, found up to 40% of transcripts to be differentially expressed. Furthermore, we discovered more than 2500 alternative splicing events, many of which appear to be stage-regulated. Based on our findings we hypothesize that alternatively spliced transcripts present a new means of regulating gene expression and could potentially contribute to protein diversity in the parasite. The entire dataset can be accessed online at TriTrypDB or through: http://splicer.unibe.ch/.

**Funding:** The work was funded through SNF grant 31003A-125194 to TO and HHMI grant 55005528 and SNF: 31003A-112092 and 31003A-126020 to IR. The funders had no role in study design, data collection and analysis, decision to publish, or preparation of the manuscript.

**Competing Interests:** The authors have declared that no competing interests exist.

* E-mail: torsten.ochsenreiter@izb.unibe.ch (TO); isabel.roditi@izb.unibe.ch (IR)

## Introduction

*Trypanosoma brucei* is a unicellular eukaryotic parasite with a digenetic life cycle alternating between the tsetse fly and a variety of mammalian hosts. Besides its importance as a human and veterinary pathogen it has been key to the discovery and understanding of general biological principles such as RNA editing, antigenic variation, GPI anchoring and *trans*-splicing [1,2,3,4,5]. The genome sequence of the 11 megabase-sized chromosomes of *T. brucei* revealed a compact structure containing about 9000 predicted genes, including 900 pseudogenes and 1700 genes specific to *T. brucei* [6]. The majority of protein coding genes in trypanosomes is organized in polycistronic units that are transcribed by RNA polymerase II (Pol II) [7]. Polycistronic RNA precursors are processed into mature monocistronic mRNAs by *trans*-splicing of a 39 nt leader sequence to the 5′ end and polyadenylation of the 3′ end [5]. With the exception of the actin promoter, no Pol II promoters for protein-coding genes have been identified, reviewed in [8]. Recently, an elegant study showed correlation between the position of histone marks and putative transcription start sites, suggesting that chromatin structure plays a major role in directing Pol II to its promoter sites [9]. While promoter structures are still elusive, it has long been known that transcription itself is regulated very little, if at all. The regulation of gene expression occurs mainly at the level of RNA stability, translation and protein stability (for review see [8]). Four microarray analyses have shown the transcriptome of the organism to be rather static, with only 2–10% of the transcripts being stage-regulated [10,11,12,13]. This limited degree of regulation at the level of transcript abundance is surprising, given the fundamental changes in morphology and metabolism that occur during the development of the parasite, and especially in the light of the large differences that occur in the closely related species *Trypanosoma cruzi*, where approximately half of the genome is regulated at the level of transcript abundance [14]. To date, very little is known about the way in which trypanosomes regulate translation, but in many other eukaryotes the 5′ untranslated regions (UTRs) contain sequence elements that are responsible for regulating protein synthesis. In order to analyze the *T. brucei* genome for such elements it is crucial to delineate the 5′ UTR of each expressed gene. In the past, bioinformatics approaches have been used to predict 5′ splice sites in *T. brucei*, but few of these have been confirmed experimentally [15]. Using a novel high throughput parallel sequencing approach that we term spliced leader trapping (SLT), we have now mapped the vast majority of 5′ UTRs and analyzed the developmental regulation of transcript abundance in bloodstream and insect form trypanosomes. SLT also provides a means of selectively analyzing the transcriptome of parasites without having to purify them from host tissue. Furthermore, since *trans*-splicing of a spliced leader has been identified not only in protozoa but also in cnidarians, nematodes, flatworms and ascidians, SLT could potentially be applied to a wide variety of organisms [5,16,17,18,19,20].

### Author Summary

Some organisms like the human and animal parasite *Trypanosoma brucei* add a leader sequence to their mRNAs through a reaction called *trans*-splicing. Until now the splice sites for most mRNAs were unknown in *T. brucei*. Using high throughput sequencing we have developed a method to identify the splice sites and at the same time measure the abundance of the corresponding mRNAs. Analyzing three different life cycle stages of the parasite we identified the vast majority of splice sites in the organism and, to our great surprise, uncovered more than 2500 alternative splicing events, many of which appeared to be specific for one of the life cycle stages. Alternative splicing is a result of the addition of the leader sequence to different positions on the mRNA, leading to mixed mRNA populations that can encode for proteins with varying properties. One of the most obvious changes caused by alternative splicing is the gain or loss of targeting signals, leading to differential localization of the corresponding proteins. Based on our findings we hypothesize that alternative splicing is a major mechanism to regulate gene expression in *T. brucei* and could contribute to protein diversity in the parasite.

## Results

### Libraries, sequencing and mapping of splice acceptor sites

Long slender and short stumpy bloodstream form trypanosomes are the proliferative and quiescent life cycle stages, respectively, in the mammalian host, while the procyclic form is a proliferative form in the midgut of the insect host. In order to map the 5′splice sites and quantify abundance of the corresponding transcripts in these three life cycle stages, poly(A)-RNA was purified and first strand cDNA was synthesized using random hexamers (Figure S1). In order to process multiple samples in one sequencing channel a four-nucleotide barcode was added to the 3′end of the cDNAs. After amplification and size fractionation to 120–160 base pairs the cDNA library was sequenced on a Genome Analyzer (GA-II, Illumina) using the Chrysalis 36 cycles v 3.0 sequencing kit and 76 cycles. Base calling was performed using the Genome Analyzer Pipeline and linker sequences were removed while separating reads according to identified barcodes. The estimated error probability of the retained bases was 1.1%. Inserts containing a barcode were up to 70 base pairs (bp) with an average insert size of 42 bp (median 45 bp, σ 17 bp; Figure 1).

We obtained 4.6 million sequence tags that could be attributed to one of the libraries (Table 1). Of these 4.5 million tags (98%) could be aligned to one of the 11 megabase-sized chromosomes of *T. brucei*, the Antat 1.1 surface glycoprotein (VSG; 227,063 reads) or the telomeric expression sites (40,333 reads). Depending on the library, 77% to 79% of all genes had at least one tag associated with their 5′ UTR or annotated coding sequence. The median number of tags per gene ranged from 12 in the long slender bloodstream form to 24 in the procyclic form (Table 1). The dynamic range of the SLT method is best described in the bloodstream library where it spanned more than 5 orders of magnitude from 1 tag for an individual gene to 205,410 tags or 11% of all tags in that library for the Antat1.1 VSG. The majority of genes without any tags were hypothetical unlikely (316), hypothetical (118), expression site associated genes (116) and hypothetical conserved genes (103).

Using information for the positions of snoRNAs, rRNAs and tRNAs, together with the direction of transcription, we estimated the number of polycistronic transcription units to be approximately 200. Transcription profile analysis of the majority of individual transcription units containing five or more protein coding genes (excluding VSG genes) indicated large variations in transcript abundance between adjacent genes. Of 270 monocistronic mRNA transcription units that have been annotated, 76 had major splice sites less than 2kb upstream of the ATG and eight were differentially expressed.

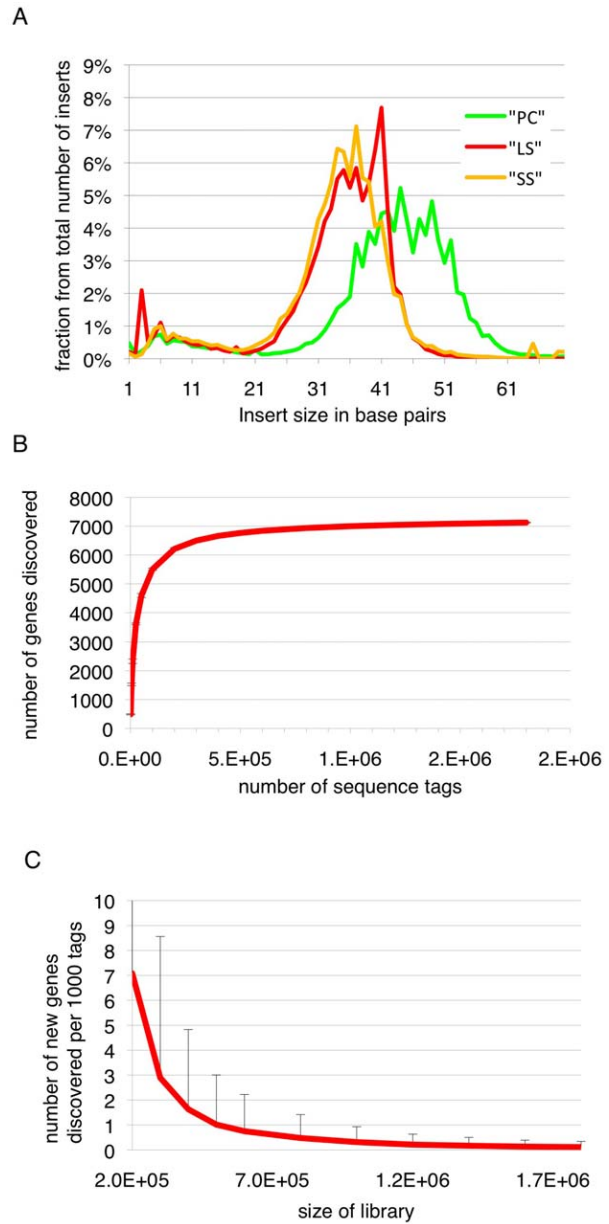

**Figure 1. Insert sizes and gene discovery rate.** (**A**) Insert sizes of the sequence tags of the three libraries in base pairs (bp); long slender bloodstream form (LS, red), short stumpy bloodstream form (SS, orange) and procyclic form (PC, green). (**B**) Simulation of gene coverage from the long slender bloodstream form library; considered are all tags that lead to UTRs ≤2 kb. Error bars depict 99.99% confidence interval from five simulations assuming normal distribution. (**C**) Number of new genes discovered per 1000 tags depending on the total number of tags, error bars depict 95% confidence interval from five simulations, assuming normal distribution.
doi:10.1371/journal.ppat.1001037.g001

**Table 1.** Library statistics.

| | Long slender bloodstream form | Short stumpy bloodstream form | Procyclic form |
|---|---|---|---|
| Number of reads | 2,223,504 | 1,919,710 | 1,430,649 |
| Number of reads ≥24nt | 1,814,248 | 1,602,918 | 1,171,012 |
| Number of reads aligned[1] | 1,562,364 | 1,538,057 | 1,148,155 |
| Number of splice sites | 29,406 | 27,572 | 23,842 |
| Genes with tags | 8,277 | 8,173 | 7,926 |
| Coverage in %[2] | 91 | 90 | 87 |
| Genes with major internal sites | 793 | 870 | 683 |
| Genes with major internal sites but no downstream AUG | 97 | 95 | 90 |
| Genes with only internal sites | 542 | 558 | 496 |
| Genes with 5′ UTR >2kb | 198 | 191 | 153 |
| Genes with major 5′ UTR >2kb | 356 | 329 | 235 |
| Genes with major 5′ UTR <2kb | 7,143 | 6,989 | 7,009 |
| Genes with ≥1 uORF | 1,331 | 1,326 | 1,229 |
| Splice sites/gene mean | 2.9 | 2.8 | 2.7 |
| Splice sites/gene median | 2 | 2 | 2 |
| 5′ UTR length mean (0–2kb) | 139 | 127 | 105 |
| 5′ UTR length median (0–2kb) | 47 | 32 | 34 |
| Tags per gene mean | 85.1 | 94.8 | 96.9 |
| Tags per gene median | 12 | 15 | 24 |

[1]reads aligned to one of the 11 megabase-sized chromosomes excluding expression sites or VSGs.
[2]of 9068 genes in the genome.
doi:10.1371/journal.ppat.1001037.t001

It has recently been shown that procyclin-associated genes (PAGs) are located preferentially at a subset of strand-switch regions, or between polycistrons that are transcribed by different polymerases, and that a region within PAG1 could silence transcription by Pol I [21]. Based on SLT analysis, mature transcripts from genes downstream of PAGs were not observed for five of the six annotated sites. In one case transcripts were absent from the procyclic form, while the bloodstream form produced spliced transcripts downstream of two PAG-like genes (Tb11.01.6210-6220). It has been documented that overlapping sense and antisense transcription can occur at a single locus, giving rise to processed mRNAs from both strands [22]. We identified 140 unique antisense spliced leader addition sites with ≥1 tag per million (TPM), 60 of which were detected in more than one life cycle stage. Thirty-eight of the unique antisense splice leader addition sites were found on the reverse strand of a hypothetical gene at the end of a transcription unit and twenty-three were in strand-switch regions.

## Expression profiling and comparison with previous transcriptome studies

In order to evaluate if the expression profiles were consistent with previously published data we first analyzed 20 well-studied genes and found good agreement between the direction and magnitude of change in transcript abundance between life cycle stages (Table S1). These included the phosphoglycerate kinases PGKB and PGKC [23], the nuclear-encoded cytochrome oxidase complex subunits (IV–X) [24], the terminal alternative oxidase [25], invariant surface glycoproteins (ISGs; [26]) and the major surface glycoproteins, VSG and procyclins [27]. In total, 3554 genes or ~40% of the genome significantly changed expression levels in at least one of the three life cycle stages (statistical significance of <10^{-5}; [28]; Table 2). More than 2000 differentially expressed transcripts could be observed between the bloodstream and procyclic forms, while 1226 changes in expression level could be detected between the proliferative long slender and quiescent short stumpy bloodstream stages. (Figure 2; Table 2). Of the entire transcriptome 5472 transcripts (60%) did not change abundance between life cycle stages; 919 of these transcripts were represented by ≥25 TPM in all stages.

We further evaluated the SLT approach by comparing the data to a previously published tiling array study [11]. Transcripts more abundant in the procyclic form correlated with a coefficient of 0.77 or 0.93 to the SLT approach, depending on the statistics used by Koumandou *et al.* (Figure S2, [11]). Transcripts more abundant in the bloodstream form showed a correlation coefficient of 0.23 or 0.43 (Figure S3). When we compared our data to the most recent microarray study by Jensen *et al.* [12], which identified 234 transcripts are being less abundant in the procyclic than the bloodstream form, 172 showed the same direction of change while 62 did not agree with our study. In addition, from the 317 transcripts that are increased in the procyclic form, 270 are in agreement with our study. Of the 551 transcripts that were significantly changed ≥2-fold between the two life cycle stages in the study by Jensen *et al.* ~80% showed the same pattern in our study. Using the two data sets, we have compiled a list of 442 genes that show a robust change in expression pattern (Table S2).

When we analyzed the metabolic pathways as annotated in KEGG we found the majority to be down regulated in the stumpy form when compared to the long slender form (Figure S4, S5, S6, Table S3). In the glycolytic pathway for example, 8 of 11 transcripts are decreased in abundance in the short stumpy form, which is in good agreement with previous studies of the metabolic

**Table 2.** Differentially expressed and spliced transcripts.

| | Long slender/ short stumpy | Short stumpy/ procyclic | Long slender/ procyclic | Total[1] |
|---|---|---|---|---|
| Significantly regulated genes | 1226 | 2675 | 3296 | 3554, (3215)[4] |
| Upregulated[2] | 769 | 1248 | 1286 | |
| Downregulated[3] | 457 | 1427 | 2010 | |
| Differentially spliced | 158 | 415 | 458 | 676 |
| Alternatively spliced | 1523 | 1531 | 1267 | 2637 |
| Alternatively spliced ≥5 tags | 874 | 783 | 872 | 1588 |

[1]total number of regulated/spliced transcripts, non-redundant.
[2]upregulated in long slender when compared with short stumpy, short stumpy vs procyclics and long slender vs procyclics.
[3]downregulated in long slender versus short stumpy, short stumpy versus procyclics and long slender vs procyclics.
[4]significantly regulated ([28]; threshold of P<10^{-5}) and ≥2× change.
doi:10.1371/journal.ppat.1001037.t002

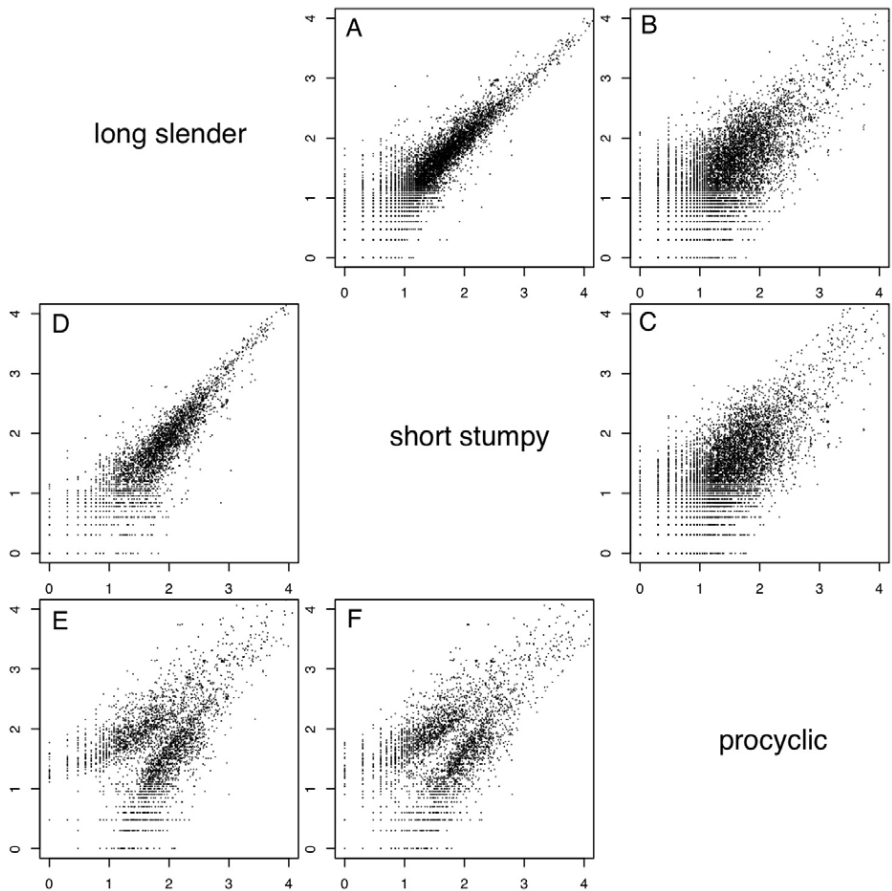

long slender

short stumpy

procyclic

**Figure 2. Whole genome comparisons of gene expression levels.** Each subfigure (**A–F**) shows a scatter plot of mRNA expression levels ($\log_{10}$ tags per million) between two life-cycle stages. Each dot represents one gene. (**A**) long slender/short stumpy, (**B**) long slender/procyclic, (**C**) short stumpy/procyclic. In panels **D–F** only genes with statistically significant difference in expression levels between any two life stages of the parasite are shown (Audic–Claverie $P < 10^5$).
doi:10.1371/journal.ppat.1001037.g002

pathways in quiescent cells in general [29] and trypanosome short stumpy form in particular (for review see [30]). We furthermore compared the expression profile of ten genes between bloodstream long slender and procyclic cells using SLT and RT qPCR and found positive correlation between the two methods (Pearson correlation r = 0.97; p-value of $p < 10^{-5}$; Figure S7, Table S4). Additionally, we performed two RNAi experiments (Alba1 and Alba 3/4) in procyclic forms (Figure S8). The efficiency of the knockdown was then measured by Northern blot analysis and SLT. For Alba1 both techniques showed a knockdown in message level to 6% compared to the uninduced cell line. For Alba 3/4 Northern blot analysis indicated a knockdown to 8% while the value measured by SLT was 13%. The overall correlation between the uninduced and the induced libraries was between 0.97 (Alba1) and 0.94 (Alba 3/4; Spearman rank correlation coefficient). Lastly we compared the expression profile from SLT with one run of regular RNAseq from poly(A)-RNA from procyclic forms (24 million sequence tags) and found a positive correlation between the two techniques (Spearman $\rho = 0.69$, Figure S9).

## Analysis of a VSG expression site

In addition to the different life cycle stages of Antat 1.1, we also analyzed the widely used monomorphic bloodstream form MITat 1.2 (221) and mapped the sequence tags to the active VSG expression site (Figure 3; [31]). The most highly expressed gene was VSG 221 with more than 69,000 TPM. The second most highly expressed gene in the expression site was the ESAG 12 gene (761 tags) followed by a hypothetical gene that has not been annotated previously, now designated ESAG13 (138 tags) and ESAG 6 and 7, which encode the two subunits of the transferrin receptor (80 and 43 tags; for review see [32]).

## Splice site detection and changes in alternative splicing patterns during development

We identified 29,406 splice sites in the *T. brucei* genome with a median of two splice sites per gene (bloodstream form AnTat1.1; Table 1). The major splice site was strictly conserved with 94% of the splice acceptor dinucleotides being AG preceded by an upstream polypyrimidine tract ($-14$ to $-43$, relative to the splice site; Figure 4A). Twenty percent of the minor splice acceptor sites contained a dinucleotide other than AG. GG occurred in 7% of these while TG, AA, GA and AC were found in 2% of the minor splice sites (Figure 4B–C). The least abundant dinucleotide was CC. When we compared the major splice sites from this study to a previous genome-wide prediction we found that about 40% of the major sites and 6% of the minor sites had correctly been predicted by the model [15]. Using the position of the major splice sites we determined the mean length of all 5′ UTRs to be 104–138 nucleotides and the median length to be 32–47 nucleotides (Figure 4D–E, Table 1). In the procyclic life cycle stage we identified 588 transcripts with splice sites that would allow N-terminal extension of the annotated protein coding gene and, for 11 of these, peptides corresponding to the N-terminal region were recently detected by mass spectrometry (Table S5; [33]). About 500 of the currently annotated genes only had internal splice sites 3′ of the predicted AUG start codon, and depending on the life cycle stage, 683–870 genes had major internal splice sites (Table 1). Of the transcripts with major internal sites ~90% had a downstream AUG in the same reading frame within 400 nucleotides of the splice site, leaving about 10% of the transcripts without a likely translation start. In each life cycle stage >1200 transcripts exhibited an alternative splicing pattern (Figure 5–6 Table 3). A transcript was termed alternatively spliced if the major splice site contained less than 60% of all sequence tags. We identified four different types of alternative splicing, based on the potential functional consequences (Figure 7). Splicing type A renders the transcript potentially untranslatable since the transcript does not contain an in-frame AUG downstream of the alternative splice site (Figure 7, Figure S10). Splicing type B leads to a 5′ truncation of the original open reading frame, with a potential downstream AUG start site (Figure S11). Splice Type C does not change the open reading frame, but it includes or excludes potential regulatory elements such as upstream open reading frames (uORFs) in the 5′ UTR (Figure S12). Splice type D potentially allows for the use of a novel open reading frame due to the inclusion of an AUG start codon in a different reading frame (Figure S13). We then analyzed if the major splice sites for a particular transcript changed between the life cycle stages and termed this differential splicing (Figure 6). Depending on the two life stages we compared, 158–458 differential splicing events affecting a total of 676 genes were identified (Table 2). We additionally verified the differential splicing patterns for three transcripts (Tb927.1.790, Tb927.6.4240, Tb11.02.2700) by RT qPCR and found positive correlation between the SLT and the RT qPCR data (Pearson r = 0.85; 95% confidence interval; Table S6).

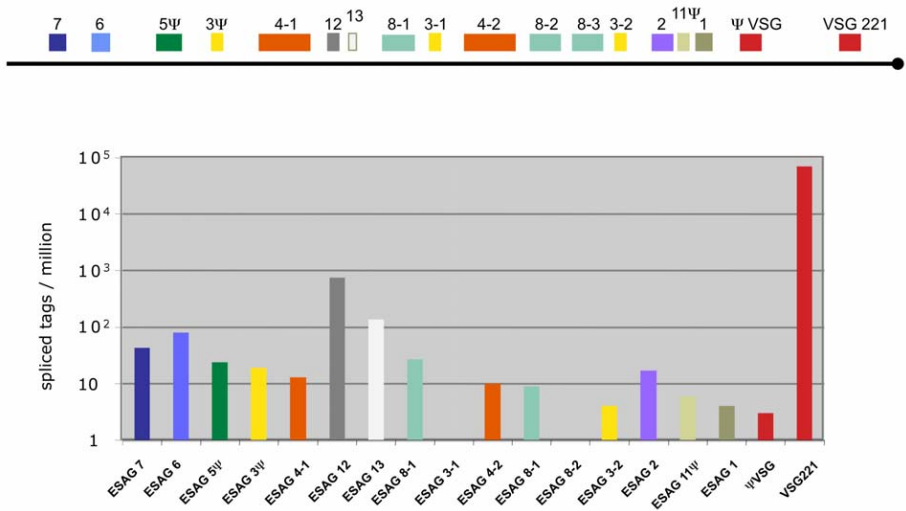

**Figure 3. Expression levels of genes in the active VSG expression site.** Schematic representation of the tag frequency distribution in the bloodstream expression site 1 (BES1/TAR40), as annotated by Hertz-Fowler *et al.* [31]. The hypothetical gene between ESAG 12 and ESAG 8-1 encodes a protein of 96 amino acids and is renamed ESAG 13.
doi:10.1371/journal.ppat.1001037.g003

## Alternative splicing of mRNAs encoding aminoacyl tRNA synthetases (AARS)

Of the 23 currently annotated distinct AARS the Asp-RS, Lys-RS, Trp-RS and Tyr-RS are each encoded by two genes, one of which contains a mitochondrial targeting signal (MTS) at the N-terminus as determined by MITOPROT (Table S7; [34]). Recently, Charrière and coworkers verified the cytosolic and mitochondrial localization of the Trp-RS and Asp-RS experimentally [35,36]. Five additional AARS (Asn-RS, Pro-RS, Glu-RS, Gln-RS, Ser-RS) contain an N-terminal MTS; for these AARS, however, there is no cytosolic isoform encoded in the genome. When we analyzed their expression and splicing patterns we found that the major 5′ splice site is internal to the currently annotated AUG start codon leaving only a small fraction of the splice sites upstream of the first AUG (Figure 8, Table S7). Thus the major translation product would exclude the MTS. Additionally, 4 AARS (Arg-RS, Cys-RS, His-RS and Leu-RS) contain an internal MTS that is masked by 48–101 amino acids at the N-terminus (Table S7). However these MTS could become N-terminal if a downstream start codon would be used as the translation start site.

## Discussion

We have developed a cost effective method to analyze whole genome expression and splicing profiles from organisms employing leader *trans*-splicing. This was tested on *T. brucei*, where we obtained 85% coverage of all genes with fewer than 1 million sequence tags. A major improvement over traditional microarray technology is the possibility of sequencing directly from the spliced leader, which allows selective analysis of the transcriptome of intracellular parasites like the amastigote forms of *Trypanosoma cruzi* or forms that are closely associated with their host and extremely difficult to purify like the epimastigote form of *T. brucei* in tsetse salivary glands.

Using the SLT approach we sequenced more than 1 million splice site tags from poly(A)-mRNA from each of three life cycle stages of *T. brucei*. Each sequence tag covered at least 24 nucleotides 3′ of the spliced leader/5′ UTR junction. Even though the analyzed strains (Antat 1.1 and Mitat 1.2) were not identical to the genome strain (TREU 927) mapping of the 24mer sequence tags onto the TREU 927 genome was very successful, given that the majority (98%) could be aligned with high statistical significance using a combination of mismatch and sequence quality scores (see materials and methods). In the process, this method revealed the existence of a new expression site associated gene in an active VSG expression site.

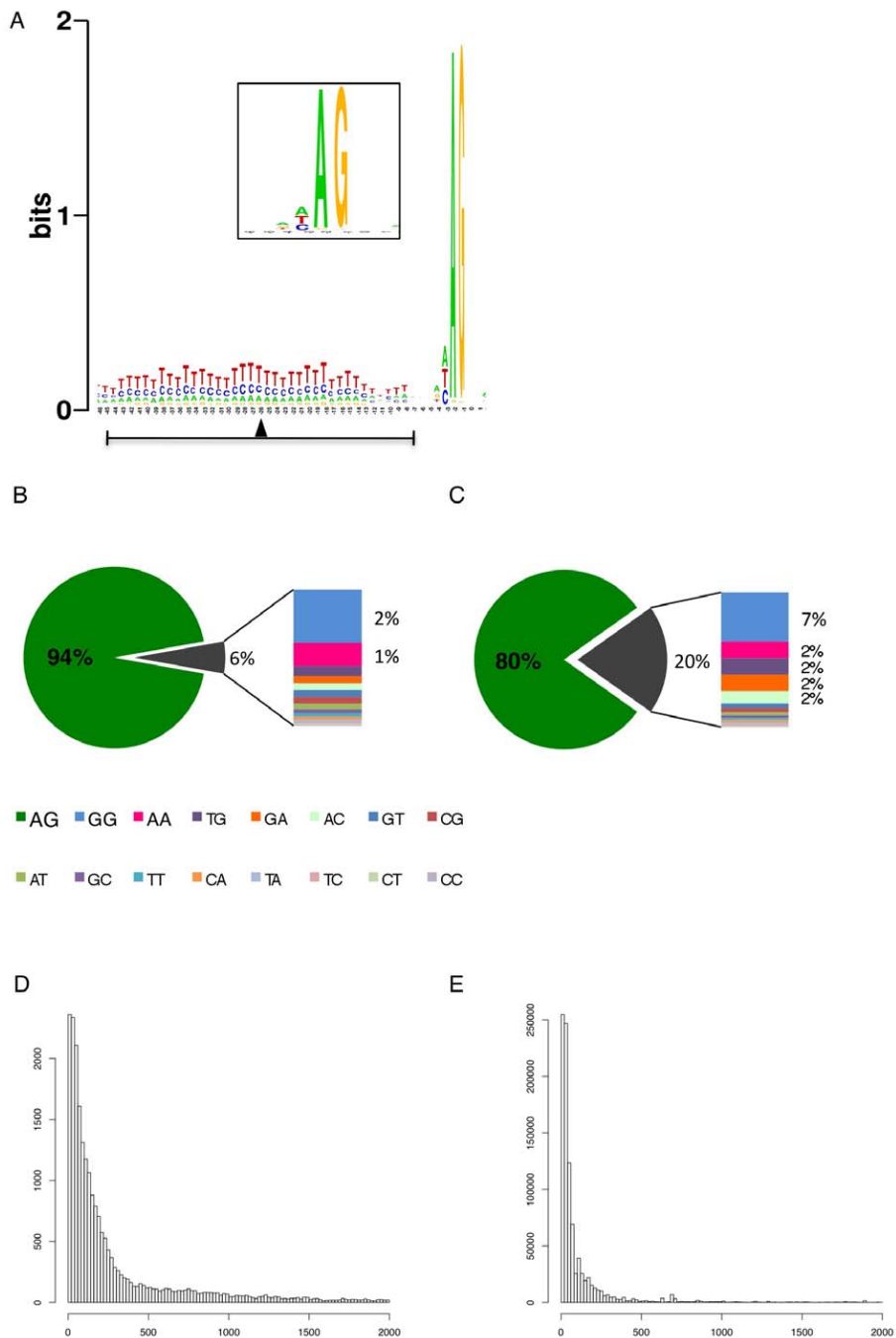

**Figure 4. Splice site conservation and 5′ UTR length distribution.** (**A**) Consensus of the major splice sites from procyclic form *T. brucei*. The inset shows a close up of the conserved AG at the splice site. The upstream polypyrimidine tract is dominated by T residues starting at an average distance of 26 nucleotides (black triangle; ±19) upstream of the splice site. (**B–C**) Splice site comparison. Abundance of dinucleotides at the splice acceptor sites for (**B**) major splice sites and (**C**) minor splice sites in *T. brucei* procyclic forms. (**D–E**) Untranslated region (5′ UTR) length distribution in procyclic forms using the major splice site on a per gene basis (**D**). 5′ UTR length distribution of the major splice site on a per transcript basis (**E**).
doi:10.1371/journal.ppat.1001037.g004

In general, our data are in good agreement with the current genome annotation, and strongly support both the number and positions of putative transcription start sites identified by binding of specific histones [8]. In addition, 98% of the sequence tags from the procyclic form map in the sense orientation of annotated transcription units, and only 2% in the zones between transcription units, where the direction of transcription is not immediately obvious from the annotations. The splicing patterns indicated that transcription overlaps in several of the converging strand switch regions as has been suggested previously for converging transcription units transcribed by Pol I and Pol II, respectively [22]. Interestingly, bloodstream form cells show twice the number of sequence tags mapping to the zones between the transcription units compared to the procyclic form, indicating that the control of transcription initiation may be less specific, or degradation less efficient in this life stage. It is also possible that splice site recognition is different, or not equally efficient between the stages, leading to the increased number of processed RNAs from the regions between the transcription units.

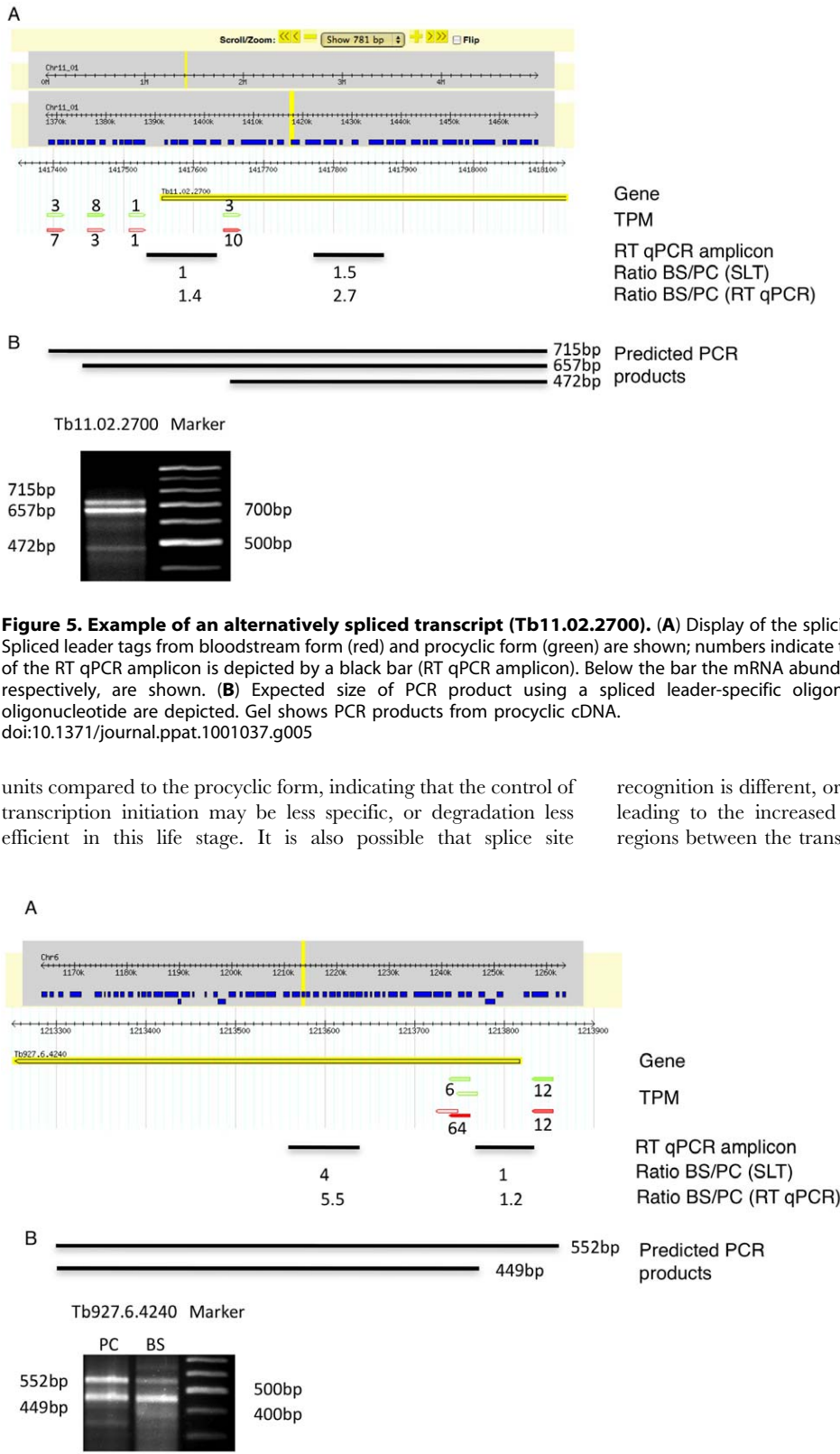

**Figure 5. Example of an alternatively spliced transcript (Tb11.02.2700).** (**A**) Display of the splicing profile for Tb11.02.2700 using Gbrowse. Spliced leader tags from bloodstream form (red) and procyclic form (green) are shown; numbers indicate tags per million (TPM). The position and size of the RT qPCR amplicon is depicted by a black bar (RT qPCR amplicon). Below the bar the mRNA abundance ratios as determined by SLT and qPCR respectively, are shown. (**B**) Expected size of PCR product using a spliced leader-specific oligonucleotide together with a gene specific oligonucleotide are depicted. Gel shows PCR products from procyclic cDNA.
doi:10.1371/journal.ppat.1001037.g005

**Figure 6. Example of a differentially spliced transcript (Tb927.6.4240).** (**A**) Display of the splicing profile for Tb927.6.4240 using Gbrowse. Spliced leader tags from bloodstream form (red) and procyclic form (green) are shown. Numbers indicate tags per million (TPM). The position and size of the RT qPCR amplicon is depicted by a black bar (RT qPCR amplicon). Below the bar the ratios of mRNA abundance as determined by SLT and qPCR respectively are shown. (**B**) Expected size of PCR products using a spliced leader-specific oligonucleotide together with a gene specific oligonucleotide are depicted. Gel shows PCR products from procyclic (PC) and bloodstream (BS, long slender) cDNA.
doi:10.1371/journal.ppat.1001037.g006

**Table 3.** Alternatively spliced transcripts.

| | Long slender | Short stumpy | Procyclic | Total[1] |
|---|---|---|---|---|
| Alternatively spliced | 1523 | 1531 | 1267 | 2637 |
| Alternatively spliced ≥5 tags | 874 | 783 | 872 | 1588 |

[1]Total number of regulated/spliced transcripts, non-redundant.
doi:10.1371/journal.ppat.1001037.t003

The large variation in transcript abundance that we detected within transcription units was expected and is in good agreement with previous studies, strengthening the notion that steady state RNA levels in trypanosomes are regulated mainly post-transcriptionally at the level of RNA processing and/or RNA stability. When we analyzed the monocistronic transcription units we found >70% unlikely to be expressed in the life cycle stages analyzed because we could not detect any splice sites within 2 kb upstream of the start codon. Of the remaining 76 transcripts, only 8 showed differential expression.

Previous microarray studies of *T. brucei* using genomic arrays or single probe arrays have indicated a rather static transcriptome with relatively few changes between the bloodstream and procyclic forms. Estimates ranged from 2–6% of the genome being regulated at the transcript abundance level [9,10,11,12,13]. A more recent study by Jensen *et al.* using eight oligonucleotides per gene on a Nimblegen microarray found that up to 700 transcripts or (8%) change expression between the two life cycle stages [12]. Considering these studies it would appear that a multiprobe array is more likely to detect a larger set of regulated genes. Furthermore, using arrays with only one probe in the 5′ region of the genes is likely to be affected by the misannotation of the start sites of open reading frames and/or by alternative splicing. Using SLT we found 30% of the transcripts from protein-coding genes to be significantly regulated between the long slender bloodstream form and the procyclic form (Table 2). The number of changes increased to 40% of all genes when extended to include the short stumpy form of the parasite. Even at a more conservative threshold (≥2-fold, significantly changed), 35% of all genes exhibited changes in transcript abundance. In conclusion, we have found that a much larger cohort of genes changes expression levels during development than was previously thought to be the case. This is in line with recent findings that about 50% of the genome of *T. cruzi* is regulated at the level of transcript abundance [14]. When we compared our results with the previous microarray studies we find different levels of correlation in gene expression depending on the study. While some part of the differences might be due to the higher sensitivity of the SLT approach we have to keep in mind that the low level of correlation might also be due to strains and culture conditions that vary considerably between the different studies. We also employed three different approaches to verify the results obtained by SLT. (i) Results from RT qPCR showed strong positive correlation with the SLT approach for the expression level changes between life stages and the differential splicing events (Figure S7, Table S4, Table S6). (ii) We also performed two RNAi experiments indicating that the abundance of the RNAi target transcript as quantified by SLT is in excellent agreement with quantification by Northern blots. The overall correlation between the induced and uninduced transcriptomes was on a par with the technical reproducibility of RNAseq, further supporting that SLT tag counts serve as a sufficient proxy for comparisons between different life cycle stages or cell lines of the same organism (Figure S8). (iii) We included the correlation to one run of regular RNAseq of poly(A)-mRNA (24 million sequence tags, Figure S9). While not perfect, the correlation between SLT and RNAseq (Spearman $\rho = 0.69$) is nearly on a par with that between a technical comparison of RNAseq and microarrays (Spearman $\rho = 0.75$ e.g. [37]).

During the revision of this manuscript Siegel *et al.* published a study describing the expression profile in bloodstream and procyclic forms of *T. brucei* using RNAseq [38]. Although the approaches used in both studies are different (RNAseq vs. SLT) many features found in both studies are well correlated; the mean number of splice sites per gene (2.6 versus 2.7–2.9), the mean lengths of the 5′ UTRs (184 versus 105–135), the number of genes with internal splice sites (488 versus 496–558) and the large dynamic range ($10^5$ to $10^6$). However, there are also a number of features that do not correlate so well, most striking of which is the

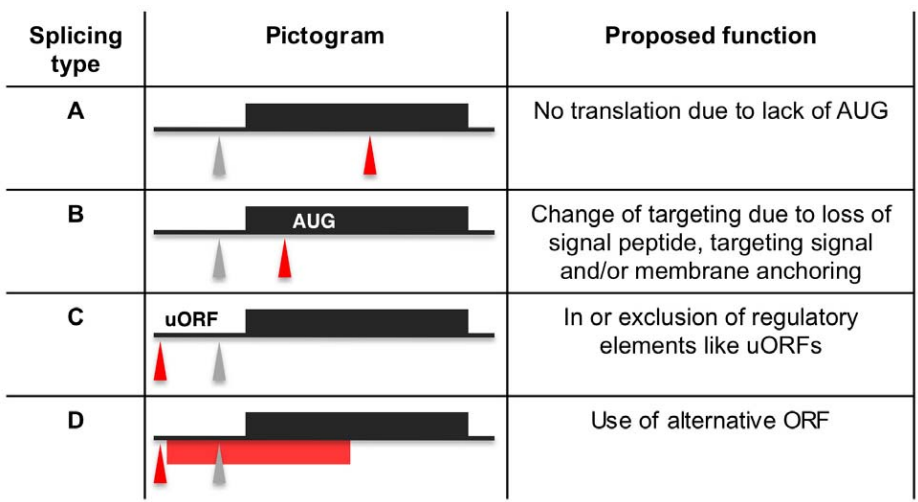

**Figure 7. Proposed functions for alternative splicing variants.** Regular splice sites (grey triangles) and alternative splice sites (red triangles) are depicted on a pre-processed mRNA (black line) containing an open reading frame (ORF; black box). The red bar depicts an alternative ORF. AUG depicts an alternative translation start site. Upstream open reading frames are denoted uORF.
doi:10.1371/journal.ppat.1001037.g007

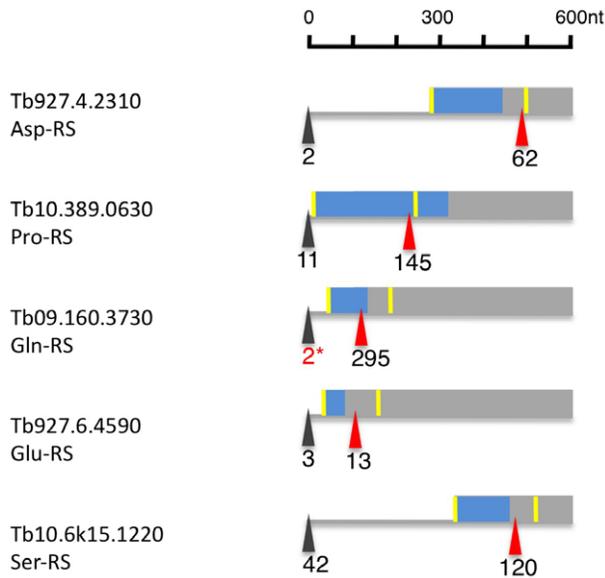

**Figure 8. Alternatively spliced aminoacyl tRNA synthetase mRNAs (AARS).** Depicted are 600 nucleotides from the 5′ region of six alternatively spliced AARS transcripts. The major spliced site (red triangle) and minor splice site (black triangle) are indicated in tags per million in procyclic forms. The yellow bar indicates the AUG downstream of the minor and major splice sites. The blue box represents the N-terminal mitochondrial targeting signal as determined by MITOPROT. The 2* marks tags, that were only seen in larger procyclic libraries (e.g. Alba1 and 3/4).

doi:10.1371/journal.ppat.1001037.g008

difference in expression profile. We identified almost 40% of the genes as being regulated significantly while Siegel *et al.* found only about 10%. There are several reasons that could explain the differences observed: (i) strain differences and the number of life stages analyzed, (ii) growth conditions for bloodstream forms (in vitro versus in vivo), (iii) cDNA preparation, with RNAseq being more likely to capture precursors and breakdown intermediates (e.g. intron sequences), and (iv) scaling of the data (tags/million versus constant median count/gene). The major advantage of SLT when compared to conventional RNAseq is the cost effective mapping of 5′ splice sites in splice leader bearing organisms, especially intracellular parasites where the purification of RNA is very difficult.

According to our study, VSG transcripts account for 7–11% of spliced mRNAs in bloodstream forms trypanosomes, which is in excellent agreement with the value obtained from hybridization data [39]. The levels of most other transcripts, however, are at least two to three orders of magnitude lower. Assuming there to be approximately 40,000 mRNA molecules in a procyclic trypanosome, and approximately half that number in a bloodstream form (Haanstra and co-workers and our calculations), a large number (>5000, 65%) would be present at <1 mRNA per cell [40]. Similar results have been reported for yeast where more than 80% of all transcripts are present at ≤2 copies per cell [41,42]. The question if the small number of transcripts is distributed evenly across the population of trypanosomes, or accumulates in very few cells during a transcriptional burst, remains to be investigated. Recently it has been suggested that transcription in yeast is much more steady, with fewer transcriptional bursts than seem to occur in mammalian cells [43,44,45].

The data presented here describes the splice sites for 85% of the annotated genes in the *T. brucei* genome (Table 1). Spliced leader addition sites were very well conserved within and between life cycle stages. AG was the predominant splice acceptor dinucleotide of the major splice sites, however 20% of the minor splice sites used other dinucleotides, predominantly GG (Figure 4A–C). The least abundant acceptor dinucleotide was CC followed by any pyrimidine combination. The major splice sites contained a 15 (±6) nucleotide polypyrimidine stretch at an average distance of 31 nucleotides (±19) upstream of the actual splice acceptor site. Both these findings are in very good agreement with previously published experimental data on individual splice sites [46]. The UTR length distribution indicated 5′ UTRs with a median length between 32 to 47 nucleotides, which is similar to yeast (50 nt; [47]; Figure 4 D–E). Interestingly we detected a shift towards longer UTRs in the long slender bloodstream form when compared to the procyclic form. This is at least in part due to the differential use of splice sites in the two life forms and might indicate differential regulation of translation. About half of the major splice sites in the different life stages could not be predicted using our current splice site recognition model, although the majority contained the signals that conform to our current understanding of splice sites. An obvious consequence of alternative splicing would be the change of N-terminal targeting sequences as has been shown for *T. cruzi* (*LYT1*) and for alternative *cis*-splicing in other systems [48,49,50,51]. Our analysis indicated this likely to be the case for several AARS that are essential in the cytosol and mitochondrion [35,36] thus providing evidence that alternative splicing is a potential mechanism for dual localization of proteins similar to what has been reported for the LYT1 gene in *T. cruzi* [48].

More than 500 transcripts with splice sites exclusively 3′ of the annotated start site were identified by SLT. While we cannot exclude that a very small fraction is still spliced upstream of the annotated AUG, we consider it much more likely that the *bona fide* start codon is within the open reading frame. A second set of transcripts (>600) indicated the possibility of 5′ extensions to the currently annotated open reading frames, which would effect changes in the N-terminus of the corresponding protein. The evaluation of these N-terminal extensions is much more difficult and will require additional experiments. However, by screening the recently published proteomics dataset from Panigrahi and co-workers, we were able to identify 11 candidates where peptides corresponding to a region upstream of the annotated start codon are expressed in the procyclic form trypanosomes [33]. Depending on the life cycle stage, we also identified 90–97 transcripts in which alternative splicing ablated the start codon, suggesting this form of splicing plays a minor but significant role in regulating gene expression.

Most surprising was that a large number of transcripts showed differential abundance of alternative splice variants in the three life stages analyzed. A previous report indicated that *T. brucei* used different splice sites on an artificial construct, but it remained unclear if, and how frequently, this might occur in the *T. brucei* transcriptome [52]. We found more than 600 differentially spliced transcripts between the life stages, supporting the idea that alternative splicing has functional consequences for the regulation of parasite development. One open question is if the actual splicing event is regulated or the differential abundance is a result of altered stability of the RNA transcripts. So far, we have been unable to detect sequence elements in the vicinity of the alternative splice sites that would explain the differential regulation of splicing itself. It is worth noting, however, that transcripts encoding several of the core components of the spliceosome, such as SMD1, SMD3 and SMG, are themselves differentially regulated during development; this may reflect an adaptation of the splicing machinery to differences in the major splicing targets, such as the VSG and procyclin transcripts, or to the subtleties of alternative splicing. These hypotheses should now be testable on a genome-wide scale, using the SLT approach in combination with RNA knockdown of specific splicing components.

## Materials and Methods

### Cell lines

*T. brucei brucei* AnTat 1.1 and MITat 1.2 (221) were used in this study. Late procyclic forms of AnTat 1.1 were cultivated in SDM79 supplemented with 10% FBS. Bloodstream forms were grown in mice. For bloodstream forms of AnTat 1.1, mice were immunosuppressed with 260mg/kg cyclophosphoamide (Sigma) 24 hours prior to intraperitoneal injection of $10^6$ parasites. Short stumpy parasites were harvested 5 days post-infection at a density of $2–4\times10^8$/ml blood. 75–80% of the cells showed a short stumpy phenotype as determined by light microscopy of blood smears after methanol fixation. Long slender forms were harvested from untreated mice at day three post-infection at a density of $<5\times10^7$ cells/ml blood. MITat 1.2 bloodstream forms were grown in mice as described above for long slender bloodstream forms of AnTat 1.1. Parasites were purified from whole blood using DE-52 anion exchange resin equilibrated to pH 8 with a bicine glucose buffer. After purification cells were centrifuged and resuspended in TriPure RNA isolation reagent (Roche, Switzerland).

### RNA extraction and library construction

RNA was extracted using TriPure (Roche, Switzerland) according to the manufacturer.

Poly(A) mRNA was purified from 7.5 µg total RNA using Dynabeads oligo-(dT) beads according to the manufacturer (Invitrogen, USA). First strand cDNA was synthesized from poly(A) RNA using random hexamers and Superscript II reverse transcriptase (Invitrogen, USA) in a final volume of 20 µl for 1 hour at 42°C. Half of the first strand mix was used for second strand synthesis (10 µl 1st strand mix, 1 µl RNaseH, 15 minutes at 37°C). Second strand synthesis was done using 2 µl 10× Thermopol buffer (New England BioLabs), 1 µl dNTPs 10 mM, 1 µl 2nd strand primer 10 µM ([BIOT]5′- AATGATACGGC-GACCACCGAGATCTACAGTTTCTGTACTATATTG -3′), 2 units Taq polymerase (New England BioLabs, USA) and 4.5 µl H2O by incubation for 5 minutes at 50°C and then 5 minutes at 72°C. The cDNA was purified on a Qiagen MinElute column (Qiagen, USA) and eluted in 10ul TE buffer. Adapter ligation to the purified dsDNA was done using (10.0 µl DNA, 2.5 µl 10× Ligase Buffer (New England BioLabs), 10.0 µl H2O, 1.0 µl Fasteris customized bar-coded paired-end Illumina adapter, 600 units T4 DNA Ligase (New England BioLabs, USA) for 1 hour at room temperature. The ligation mix was purified from unligated linker with streptavidin beads (Dynabeads). After incubation at room temperature for 15 minutes beads were separated on a magnetic stand, and washed twice and then resuspended in 20 µl 10 mM Tris buffer. 5 µl was used for PCR amplification with primers (5′- AATGATACGGCGAC-CACCGA -3′/5′- CAAGCAGAAGACGGCATACGAGATCG-GTCTCGGCATTCCTGCTGAAC -3′) following the standard Illumina mRNA-SEQ library amplification protocol. Fragments in the range 120–160 bp were separated on a 2% agarose-TBE gel and subsequently purified on Qiagen Gel Extraction MinElute columns (Qiagen, USA). Quality control for insert size was performed by TOPO cloning and subsequent ABI sequencing. Sequencing on the Illumina Genome Analyzer was carried out using the following sequencing primer (5′- ACCGAGATCTA-CAGTTTCTGTACTATATTG -3′).

In total we sequenced three libraries from bloodstream form mRNA, (long slender and short stumpy, both Antat1.1 and monomorphic Lister 427), one from procyclic form mRNA (Antat1.1), 2×2 RNAi libraries (uninduced and induced), where each pair could be considered a biological replicate. Lastly we prepared and sequenced one conventional RNAseq library from procyclic mRNA (Antat1.1).

### Bioinformatic analysis

Base calling was performed using the Genome Analyzer Pipeline (Illumina). Linker sequences were removed while separating reads according to identified barcodes. Only sequence reads of inserts with a length of at least 24 nucleotides were retained. A pipeline was set up using languages with open source interpreters (bash, perl and R) to automate the following analysis. The reads were mapped to the genome sequence of *T. brucei* TREU927 using maq ([53] http://maq.sourceforge.net) with n = 3 and an effective first read length of 24. Single mapping reads were separated from multi mapping reads by an alignment quality threshold of 30. Tags ending in the same position were joined together, and their counts were added (using bioperl, [54]). Tag counts were normalized to the library size (number of reads of length 24 or more) and scaled linearly to reflect counts of tags per million (TPM). Mapped tags were assigned to the annotated protein coding gene 3′ of the tag. Tags mapping internally to a coding sequence (CDS) were assigned to it as internal splice sites. Data was exported in tabular and GFF format (http://www.sanger.ac.uk/Software/formats/GFF/) and then visualized using Gbrowse [55]. Alternative spliced leader addition sites were cataloged for each gene. Genes with ≤60% tags in the assigned major splice site were designated alternatively spliced. The 5′ UTR lengths were calculated and visualized using R [56]. Upstream open reading frames (uORFs) were detected on the mapped 5′ UTR, counted and assigned a note describing their length, whether they were terminated on the 5′ UTR, or overlapped the *bona fide* (CDS) start codon. Sequences surrounding the splice sites were extracted using bioperl and visualized as sequence logos using WebLogo [57,58]. Differences in expression levels of a gene in two stages was tested for significance according to Audic and Claverie with a threshold of $P<10^{-5}$. Scatterplots of the differential gene expression levels of all libraries were produced using R [28]. When a gene had different major splice sites in two stages, the normalized counts of these sites were tested for a statistically significant difference (Fisher two-tailed test, $P<10^{-5}$). Significant differences were termed differential *trans*-splicing events. Expression levels were pooled over *T. brucei* specific KEGG pathways and visualized as heatmaps after $\log_2$ transform and hierarchical average linkage clustering of euclidian distances using MeV ([59]; http://www.genome.jp/kegg; http://www.tm4.org/mev.html).

Anti-sense splice sites were detected using bioperl. In order to simulate gene coverage a subset of reads was drawn randomly without replacement from one library and the mapping and analysis pipeline executed for each subset. This was repeated five times for each subset size. Saturation curves were drawn for several parameters, with error bars given as confidence intervals assuming normal distribution. Comparison to splice model predictions were made according to Benz *et al.* with the mapped splice sites substituted for the previous small mapping EST set [15]. All bioinformatics tools, programs, pipelines used in this study will be provided upon request. All sequence data including the regular RNAseq data is available through our website and TriTrypDB.

## cDNA and PCR and RT qPCR

PCR was used to confirm the splice sites detected by SLT. 3 µg total RNA was used as a template for reverse transcription in 50 µl AMV reverse transcriptase buffer (Promega, USA) in the presence of 1mM dNTPs, 360 ng random hexamers, 80 U RNasIn (Promega, USA) and 60 U AMV reverse transcriptase (Promega, USA). Subsequently 1µl of this reaction was used for PCR. For PCR the splice leader primer CGCTATTATTAGAACA-GTTTCGTGTAC-3′ (Tm 55°C), and reverse primers 5′-GTTGCATCCGGTGTTCTTTT -3′ (Tb11.02.2700) and 5′-AGCAGTCATCAATTCTTCCT-3′ (Tb927.6.4240) were used. The reaction was performed in 25µl, (2mM MgCl₂, 2.5µl PCR 10× buffer, 400nM primer, 1 unit of *Taq* DNA polymerase (Qiagen, USA), 0.2mM dNTP and 30ng/ul cDNA) for 30 cycles. RT qPCR was done essentially as described previously. cDNA for RT-qPCR was prepared as described above. The primers were designed such they would amplify regions of 80 to 150 nucleotides using the online software tool for real time PCR from Genescript (Table S8). Real-time PCR was run on the GeneAmp 7000 (Applied Biosystems) using 30ng of cDNA, 400nM oligonucleotides in 25µl of the MESA GREEN qPCR Master mix for SYBR assay (Eurogentec). Values were normalized to beta tubulin and the amplification efficiency was derived from a cDNA dilution series covering five logs. Average values and standard deviations of 3 RT qPCRs from one cDNA sample are shown.

## Sequence data

The sequence data can be accessed through GEO (http://www.ncbi.nlm.nih.gov/geo/) with the accession number GSE22571 or through our website http://splicer.unibe.ch/ or in future also from TriTrypDB (http://tritrypdb.org/tritrypdb/).

## Ethics statement

All animal work has been conducted in accordance with the regulations for the production of bloodstream form trypanosomes of the cantonal animal protection agency in Bern, Switzerland. Approved by the cantonal animal protection agency in Bern Switzerland (form B, 14/07).

## Supporting Information

**Figure S1** Spliced leader trapping approach. Schematic description of (A) library preparation, sequencing and (B) bioinformatics analysis.
Found at: doi:10.1371/journal.ppat.1001037.s001 (0.62 MB PDF)

**Figure S2** Microarray correlation. Correlation between the SLT approach and a recent microarray study by Koumandou *et al.* (2008). Transcripts more abundant in procyclic forms than in the bloodstream forms correlated with a coefficient of 0.77 or 0.93 to the SLT approach depending on the statistics (fspma or limma) used in the study by Koumandou and coworkers [11].
Found at: doi:10.1371/journal.ppat.1001037.s002 (0.66 MB PDF)

**Figure S3** Microarray correlation. Correlation between the SLT approach and a recent microarray study by Koumandou *et al.* (2008). Transcripts more abundant in bloodstream forms than the procyclic forms correlated with a coefficient of 0.43 or 0.23 to the SLT approach depending on the statistics (fspma or limma) used in the study by Koumandou and coworkers [11].
Found at: doi:10.1371/journal.ppat.1001037.s003 (0.38 MB PDF)

**Figure S4** Expression profile of KEEG pathways. Heatmap of log₂ changes for KEGG pathway genes for three life cycle stages of *T. brucei*. The dendrogram was obtained using hierarchical average linkage clustering of euclidian distances. LS, long slender bloodstream form; SS: short stumpy bloodstream form; PC, procyclic form.
Found at: doi:10.1371/journal.ppat.1001037.s004 (0.18 MB PDF)

**Figure S5** Expression profile glycolysis pathway. Regulation of the glycolytic pathway as annotated in KEGG. The differences in expression between long slender bloodstream and short stumpy (maroon), short stumpy and procyclic (light green) and long slender bloodstream from and procyclic form (dark green) are shown in log₂ fold.
Found at: doi:10.1371/journal.ppat.1001037.s005 (0.09 MB PDF)

**Figure S6** Expression profile oxidative phosphorylation pathway. The regulation of the citric acid cycle pathway as annotated in KEGG. The differences in expression between long slender bloodstream and short stumpy (maroon), short stumpy and procyclic (light green) and long slender bloodstream form and procyclic form (dark green) are shown in log₂ fold.
Found at: doi:10.1371/journal.ppat.1001037.s006 (0.26 MB PDF)

**Figure S7** Correlation RT qPCR and SLT. Comparison of expression levels between SLT and RT qPCR from procyclic over long slender bloodstream form cells. Log₂ fold changes of expression are shown for the SLT (blue) and RT qPCR. A positive value indicates higher steady state level RNA abundance for the corresponding gene in procyclic cells.
Found at: doi:10.1371/journal.ppat.1001037.s007 (0.17 MB PDF)

**Figure S8** Comparison RNAi libraries uninduced induced. Comparison of SLT and Northern hybridization. (A and B) Relative levels of RNA transcripts detected by either SLT or Northern blot analysis using RNAi cell lines Alba1 and Alba 3/4. −Tet/+Tet refers to RNA isolated from uninduced and induced cell lines after 7 days. Quantitation of Northern blots was performed with a Phosphoimager using the 18S rRNA as a loading control. RNA levels from uninduced cells were set at 100%. (C and D) Scatter plots of mRNA expression levels (log10 tags per million) between two libraries. Each dot represents one gene. (C) RNAi experiment Alba1_N non-induced vs. Alba1_I induced. (D) RNAi experiment Alba34_N non-induced vs. Alba34_I induced. (E) Overall correlation of steady state RNA levels of the entire genome from uninduced and induced cell lines.
Found at: doi:10.1371/journal.ppat.1001037.s008 (0.27 MB PDF)

**Figure S9** Scatter plot expression profile comparison SLT and RNA seq. Scatter plot of mRNA expression levels (log₁₀ tags per million) between SLT and RNAseq from poly(A) procyclic *T. brucei* mRNA.
Found at: doi:10.1371/journal.ppat.1001037.s009 (0.09 MB PDF)

**Figure S10** Alternative splice variant A. Example of splicing type A, no downstream AUG is found in the reading frame of the annotated gene. This also represents an example of a differentially spliced transcript, where the major splice site changes between long slender (LS), short stumpy (SS) and procyclic form (PC). RT qPCR amplicon shows the region that was used for RT qPCR. Ratio BS/PC (SLT) indicates the ratio of SLT tags, for the downstream tags the upstream tags are added in order to be comparable to the qPCR results (i.e. 1 downstream+9 upstream).
Found at: doi:10.1371/journal.ppat.1001037.s010 (0.07 MB PDF)

**Figure S11** Alternative splice variant B. Example of splicing type B with a downstream AUG (M) in the reading frame of the annotated gene, but use of that AUG would lead to loss of the signal peptide predicted by SingalP. This also represents an example of a differentially spliced transcript where the major splice site changes between the long slender (LS), short stumpy (SS) and the procyclic form (PC).

Found at: doi:10.1371/journal.ppat.1001037.s011 (0.08 MB PDF)

**Figure S12** Alternative splice variant C. Example of splicing type C with several uORFs between the two alternative splice sites. uORFs are in color according to the reading frame. The minimum length for an uORF was set to six amino acids. Long slender (LS), short stumpy (SS) and the procyclic form (PC).

Found at: doi:10.1371/journal.ppat.1001037.s012 (0.08 MB PDF)

**Figure S13** Alternative splice variant D. Example of splicing type D with an overlapping open reading frame of 384 bases (ORF2). Long slender (LS), short stumpy (SS) and procyclic form (PC).

Found at: doi:10.1371/journal.ppat.1001037.s013 (0.09 MB PDF)

**Table S1** Correlation of SLT expression profile
Found at: doi:10.1371/journal.ppat.1001037.s014 (0.04 MB PDF)

**Table S2** Robustly regulated transcripts
Found at: doi:10.1371/journal.ppat.1001037.s015 (0.08 MB PDF)

**Table S3** KEGG pathways and their expression levels in three life cycle stage
Found at: doi:10.1371/journal.ppat.1001037.s016 (0.08 MB PDF)

**Table S4** Correlation of expression profile between SLT and qPCR for 10 selected genes
Found at: doi:10.1371/journal.ppat.1001037.s017 (0.03 MB PDF)

**Table S5** Proteins with N-terminal extensions
Found at: doi:10.1371/journal.ppat.1001037.s018 (0.04 MB PDF)

**Table S6** Comparison of expression levels from differential splice sites using RT qPCR and SLT
Found at: doi:10.1371/journal.ppat.1001037.s019 (0.04 MB PDF)

**Table S7** Alternative splicing and dual localization of tRNA Synthetases
Found at: doi:10.1371/journal.ppat.1001037.s020 (0.07 MB PDF)

**Table S8** Oligonucleotide sequences
Found at: doi:10.1371/journal.ppat.1001037.s021 (0.04 MB PDF)

## Acknowledgments

We would like to acknowledge Astrid Chanfon for excellent technical assistance and the reviewers for critical suggestions that helped improve the manuscript.

## Author Contributions

Conceived and designed the experiments: IR TO. Performed the experiments: DN KG MO TO. Analyzed the data: DN KG LB IR TO. Contributed reagents/materials/analysis tools: JM LF TO. Wrote the paper: IR TO.

## References

1. Benne R, Van den Burg J, Brakenhoff JP, Sloof P, Van Boom JH, et al. (1986) Major transcript of the frameshifted coxII gene from trypanosome mitochondria contains four nucleotides that are not encoded in the DNA. Cell 46: 819–826.
2. Borst P, Cross GA (1982) Molecular basis for trypanosome antigenic variation. Cell 29: 291–303.
3. Ferguson MA, Cross GA (1984) Myristylation of the membrane form of a Trypanosoma brucei variant surface glycoprotein. J Biol Chem 259: 3011–3015.
4. de Almeida ML, Turner MJ, Stambuk BB, Schenkman S (1988) Identification of an acid-lipase in human serum which is capable of solubilizing glycophosphatidylinositol-anchored proteins. Biochem Biophys Res Commun 150: 476–482.
5. Sutton RE, Boothroyd JC (1986) Evidence for trans splicing in trypanosomes. Cell 47: 527–535.
6. Berriman M, Ghedin E, Hertz-Fowler C, Blandin G, Renauld H, et al. (2005) The genome of the African trypanosome Trypanosoma brucei. Science 309: 416–422.
7. Martinez-Calvillo S, Yan S, Nguyen D, Fox M, Stuart K, et al. (2003) Transcription of Leishmania major Friedlin chromosome 1 initiates in both directions within a single region. Mol Cell 11: 1291–1299.
8. Clayton CE (2002) Life without transcriptional control? From fly to man and back again. Embo J 21: 1881–1888.
9. Siegel TN, Hekstra DR, Kemp LE, Figueiredo LM, Lowell JE, et al. (2009) Four histone variants mark the boundaries of polycistronic transcription units in Trypanosoma brucei. Genes Dev 23: 1063–1076.
10. Brems S, Guilbride DL, Gundlesdodjir-Planck D, Busold C, Luu VD, et al. (2005) The transcriptomes of Trypanosoma brucei Lister 427 and TREU927 bloodstream and procyclic trypomastigotes. Mol Biochem Parasitol 139: 163–172.
11. Koumandou VL, Natesan SK, Sergeenko T, Field MC (2008) The trypanosome transcriptome is remodeled during differentiation but displays limited responsiveness within life stages. BMC Genomics 9: 298.
12. Jensen BC, Sivam D, Kifer CT, Myler PJ, Parsons M (2009) Widespread variation in transcript abundance within and across developmental stages of Trypanosoma brucei. BMC Genomics 10: 482.
13. Kabani S, Fenn K, Ross A, Ivens A, Smith TK, et al. (2009) Genome-wide expression profiling of in vivo-derived bloodstream parasite stages and dynamic analysis of mRNA alterations during synchronous differentiation in Trypanosoma brucei. BMC Genomics 10: 427.
14. Minning TA, Weatherly DB, Atwood J, 3rd, Orlando R, Tarleton RL (2009) The steady-state transcriptome of the four major life-cycle stages of Trypanosoma cruzi. BMC Genomics 10: 370.
15. Benz C, Nilsson D, Andersson B, Clayton C, Guilbride DL (2005) Messenger RNA processing sites in Trypanosoma brucei. Mol Biochem Parasitol 143: 125–134.
16. Krause M, Hirsh D (1987) A trans-spliced leader sequence on actin mRNA in C. elegans. Cell 49: 753–761.
17. Rajkovic A, Davis RE, Simonsen JN, Rottman FM (1990) A spliced leader is present on a subset of mRNAs from the human parasite Schistosoma mansoni. Proc Natl Acad Sci U S A 87: 8879–8883.
18. Tessier LH, Keller M, Chan RL, Fournier R, Weil JH, et al. (1991) Short leader sequences may be transferred from small RNAs to pre-mature mRNAs by trans-splicing in Euglena. EMBO J 10: 2621–2625.
19. Vandenberghe AE, Meedel TH, Hastings KE (2001) mRNA 5′-leader trans-splicing in the chordates. Genes Dev 15: 294–303.
20. Zhang H, Hou Y, Miranda L, Campbell DA, Sturm NR, et al. (2007) Spliced leader RNA trans-splicing in dinoflagellates. Proc Natl Acad Sci U S A 104: 4618–4623.
21. Haenni S, Studer E, Burkard GS, Roditi I (2009) Bidirectional silencing of RNA polymerase I transcription by a strand switch region in Trypanosoma brucei. Nucleic Acids Res 37: 5007–5018.
22. Liniger M, Bodenmuller K, Pays E, Gallati S, Roditi I (2001) Overlapping sense and antisense transcription units in Trypanosoma brucei. Mol Microbiol 40: 869–878.
23. Colasante C, Robles A, Li CH, Schwede A, Benz C, et al. (2007) Regulated expression of glycosomal phosphoglycerate kinase in Trypanosoma brucei. Mol Biochem Parasitol 151: 193–204.
24. Mayho M, Fenn K, Craddy P, Crosthwaite S, Matthews K (2006) Post-transcriptional control of nuclear-encoded cytochrome oxidase subunits in Trypanosoma brucei: evidence for genome-wide conservation of life-cycle stage-specific regulatory elements. Nucleic Acids Res 34: 5312–5324.
25. Chaudhuri M, Sharan R, Hill GC (2002) Trypanosome alternative oxidase is regulated post-transcriptionally at the level of RNA stability. J Eukaryot Microbiol 49: 263–269.
26. Ziegelbauer K, Overath P (1992) Identification of invariant surface glycoproteins in the bloodstream stage of Trypanosoma brucei. J Biol Chem 267: 10791–10796.
27. Roditi I, Carrington M, Turner M (1987) Expression of a polypeptide containing a dipeptide repeat is confined to the insect stage of Trypanosoma brucei. Nature 325: 272–274.
28. Audic S, Claverie JM (1997) The significance of digital gene expression profiles. Genome Res 7: 986–995.
29. Brecht M, Parsons M (1998) Changes in polysome profiles accompany trypanosome development. Mol Biochem Parasitol 97: 189–198.
30. Gray JV, Petsko GA, Johnston GC, Ringe D, Singer RA, et al. (2004) "Sleeping beauty": quiescence in Saccharomyces cerevisiae. Microbiol Mol Biol Rev 68: 187–206.
31. Hertz-Fowler C, Figueiredo LM, Quail MA, Becker M, Jackson A, et al. (2008) Telomeric expression sites are highly conserved in Trypanosoma brucei. PLoS One 3: e3527.
32. Borst P, Fairlamb AH (1998) Surface receptors and transporters of Trypanosoma brucei. Annu Rev Microbiol 52: 745–778.

33. Panigrahi AK, Ogata Y, Zikova A, Anupama A, Dalley RA, et al. (2009) A comprehensive analysis of Trypanosoma brucei mitochondrial proteome. Proteomics 9: 434–450.

34. Claros MG, Vincens P (1996) Computational method to predict mitochondrially imported proteins and their targeting sequences. Eur J Biochem 241: 779–786.

35. Charriere F, Helgadottir S, Horn EK, Soll D, Schneider A (2006) Dual targeting of a single tRNA(Trp) requires two different tryptophanyl-tRNA synthetases in Trypanosoma brucei. Proc Natl Acad Sci U S A 103: 6847–6852.

36. Charriere F, O'Donoghue P, Helgadottir S, Marechal-Drouard L, Cristodero M, et al. (2009) Dual targeting of a tRNAAsp requires two different aspartyl-tRNA synthetases in Trypanosoma brucei. J Biol Chem 284: 16210–16217.

37. Marioni JC, Mason CE, Mane SM, Stephens M, Gilad Y (2008) RNA-seq: an assessment of technical reproducibility and comparison with gene expression arrays. Genome Res 18: 1509–1517.

38. Siegel TN, Hekstra DR, Wang X, Dewell S, Cross GA Genome-wide analysis of mRNA abundance in two life-cycle stages of Trypanosoma brucei and identification of splicing and polyadenylation sites. Nucleic Acids Res.

39. De Lange T, Michels PA, Veerman HJ, Cornelissen AW, Borst P (1984) Many trypanosome messenger RNAs share a common 5′ terminal sequence. Nucleic Acids Res 12: 3777–3790.

40. Haanstra JR, Stewart M, Luu VD, van Tuijl A, Westerhoff HV, et al. (2008) Control and regulation of gene expression: quantitative analysis of the expression of phosphoglycerate kinase in bloodstream form Trypanosoma brucei. J Biol Chem 283: 2495–2507.

41. Yassour M, Kaplan T, Fraser HB, Levin JZ, Pfiffner J, et al. (2009) Ab initio construction of a eukaryotic transcriptome by massively parallel mRNA sequencing. Proc Natl Acad Sci U S A 106: 3264–3269.

42. Holstege FC, Jennings EG, Wyrick JJ, Lee TI, Hengartner CJ, et al. (1998) Dissecting the regulatory circuitry of a eukaryotic genome. Cell 95: 717–728.

43. Zenklusen D, Larson DR, Singer RH (2008) Single-RNA counting reveals alternative modes of gene expression in yeast. Nat Struct Mol Biol 15: 1263–1271.

44. Newlands S, Levitt LK, Robinson CS, Karpf AB, Hodgson VR, et al. (1998) Transcription occurs in pulses in muscle fibers. Genes Dev 12: 2748–2758.

45. Femino AM, Fay FS, Fogarty K, Singer RH (1998) Visualization of single RNA transcripts in situ. Science 280: 585–590.

46. Siegel TN, Tan KS, Cross GA (2005) Systematic study of sequence motifs for RNA trans splicing in Trypanosoma brucei. Mol Cell Biol 25: 9586–9594.

47. Vassella E, Braun R, Roditi I (1994) Control of polyadenylation and alternative splicing of transcripts from adjacent genes in a procyclin expression site: a dual role for polypyrimidine tracts in trypanosomes? Nucleic Acids Res 22: 1359–1364.

48. Benabdellah K, Gonzalez-Rey E, Gonzalez A (2007) Alternative trans-splicing of the Trypanosoma cruzi LYT1 gene transcript results in compartmental and functional switch for the encoded protein. Mol Microbiol 65: 1559–1567.

49. Lopez V, Kelleher SL (2009) Zinc transporter-2 (ZnT2) variants are localized to distinct subcellular compartments and functionally transport zinc. Biochem J 422: 43–52.

50. Juneau K, Nislow C, Davis RW (2009) Alternative splicing of PTC7 in Saccharomyces cerevisiae determines protein localization. Genetics 183: 185–194.

51. Ashibe B, Hirai T, Higashi K, Sekimizu K, Motojima K (2007) Dual subcellular localization in the endoplasmic reticulum and peroxisomes and a vital role in protecting against oxidative stress of fatty aldehyde dehydrogenase are achieved by alternative splicing. J Biol Chem 282: 20763–20773.

52. Helm JR, Wilson ME, Donelson JE (2008) Different trans RNA splicing events in bloodstream and procyclic Trypanosoma brucei. Mol Biochem Parasitol 159: 134–137.

53. Li H, Ruan J, Durbin R (2008) Mapping short DNA sequencing reads and calling variants using mapping quality scores. Genome Res 18: 1851–1858.

54. Stajich JE, Block D, Boulez K, Brenner SE, Chervitz SA, et al. (2002) The Bioperl toolkit: Perl modules for the life sciences. Genome Res 12: 1611–1618.

55. Stein LD, Mungall C, Shu S, Caudy M, Mangone M, et al. (2002) The generic genome browser: a building block for a model organism system database. Genome Res 12: 1599–1610.

56. R-Development-Core-Team (2009) R: A Language and Environment for Statistical Computing. Vienna, Austria.

57. Schneider TD, Stephens RM (1990) Sequence logos: a new way to display consensus sequences. Nucleic Acids Res 18: 6097–6100.

58. Crooks GE, Hon G, Chandonia JM, Brenner SE (2004) WebLogo: a sequence logo generator. Genome Res 14: 1188–1190.

59. Kanehisa M, Goto S (2000) KEGG: kyoto encyclopedia of genes and genomes. Nucleic Acids Res 28: 27–30.